\documentclass[aip,jcp,reprint,superscriptaddress,amsmath,amssymb,floatfix]{revtex4-2}

\usepackage[utf8]{inputenc}
\usepackage[T1]{fontenc}
\usepackage{graphicx}
\usepackage{amsmath,amssymb}
\usepackage{bm}
\usepackage{braket}
\usepackage{mhchem} % for chemical formulas
\usepackage{hyperref}
\usepackage{color}
\usepackage{comment}
\usepackage[normalem]{ulem} % for strikeout text
\usepackage{multirow}
\usepackage{dcolumn} % Align table columns on decimal point
\usepackage{physics} % Optional: for bra-ket and operators
\usepackage{siunitx} % For units
\usepackage{cleveref}
\usepackage{subfigure}

\hypersetup{
    colorlinks=true,
    linkcolor=blue,
    citecolor=blue,
    urlcolor=blue,
    pdftitle={Mode-Selective Spin Relaxation in a Molecular Qubit},
    pdfauthor={Your Name et al.}
}

\usepackage{xr-hyper}

\begin{document}

% % LA-UR-24-27914
% Roman Dmitriev
% %ORCID: 0009-0007-3522-7204
% Nosheen Younas
% ORCID: 0009-0006-3375-9947
% Yu Zhang
% ORCID:  0000-0001-8938-1927
% Andrei Piryatinski
%ORCID:  0000-0001-9218-1678
% Eric R. Bittner
%ORCID: 0000-0002-0775-9664
% 

\title{Tensorial Spin–Phonon Relaxation Reveals Mode-Selective Relaxation Pathways in a Single-Molecule Magnet}

\author{Roman Dmitriev}
\affiliation{Department of Physics, University of Houston, Houston, Texas 77204, USA}

\author{Nosheen Younas}
\affiliation{Department of Physics, University of Houston, Houston, Texas 77204, USA}
\affiliation{Theoretical Division, Los Alamos National Laboratory, Los Alamos, New Mexico 87545, USA}
\affiliation{Center for Nonlinear Studies, Los Alamos National Laboratory, Los Alamos, New Mexico 87545, USA}

\author{Yu Zhang}
\affiliation{Theoretical Division, Los Alamos National Laboratory, Los Alamos, New Mexico 87545, USA}

\author{Andrei Piryatinski}
\affiliation{Theoretical Division, Los Alamos National Laboratory, Los Alamos, New Mexico 87545, USA}

\author{Eric R. Bittner}
\email{ebittner@central.uh.edu}
\affiliation{Department of Physics, University of Houston, Houston, Texas 77204, USA}

\begin{abstract}
Understanding and controlling spin relaxation in molecular qubits is essential for developing chemically tunable quantum information platforms. We present a fully first-principles framework for computing the spin relaxation tensor in a single-molecule magnet, \ce{VOPc(OH)8}, by combining density functional theory with a mode-resolved open-system formalism. By expanding the spin Hamiltonian in vibrational normal modes and evaluating both linear and quadratic spin–phonon coupling tensors via finite differences of the $g$-tensor, we construct a relaxation tensor that enters a Lindblad-type quantum master equation. Our formalism captures both direct (one-phonon) and resonant-Raman (two-phonon) relaxation processes. Numerical analysis reveals a highly mode-selective structure: only three vibrational modes dominate longitudinal ($T_1$) decoherence, while a single mode accounts for the majority of transverse ($T_2$) relaxation. The computed relaxation times show excellent agreement with experimental measurements, without any empirical fitting. These results demonstrate that first-principles spin–phonon tensors can provide predictive insight into decoherence pathways and guide the rational design of molecular qubits.
\end{abstract}
%%%%%%%%%%%%%%%%%%%%%%%%%%%

\maketitle

\section{Introduction}
Recent efforts have explored various physical systems for qubit architectures in the pursuit of practical quantum computing realisation, including superconducting circuits~\cite{Krantz2019}, trapped ions~\cite{Blatt:2012ux}, quantum dots~\cite{Loss1998}, single-molecular magnets (SMMs)~\cite{Bertaina:2008wt}, and neutral atoms~\cite{Lukin:2001uo}. 
Molecular magnets and metal–organic frameworks (MOFs) have recently attracted attention for their potential in quantum information processing, high-density magnetic data storage, and as magnetic resonance contrast agents~\cite{GaitaArino2019, wasielewski2020exploiting, Iqbal:2024aa, Yu:2020uo, Yamabayashi:2018uh, Jellen:2020tx}. 
These materials offer the ability to tailor magnetic behaviour through the synthetic versatility of molecular compounds. 
However, they face significant challenges due to their susceptibility to molecular vibrations, which can strongly impact qubit performance.
The interaction between spins and molecular vibrations plays a critical role in molecular imaging, optoelectronics, and quantum technology, influencing applications such as MRI, energy harvesting, and the understanding of decoherence~\cite{wasielewski2020exploiting}. 
This spin–phonon coupling is particularly significant in SMMs, which are known for their magnetic bistability and memory retention at low temperatures. 
MOF qubit engineering therefore aims to slow spin relaxation dynamics in order to enhance magnetic memory and quantum coherence.

The primary challenge faced by these systems lies in their vulnerability to environmental decoherence, particularly through coupling between localized electronic spins and the vibrational modes of the molecular scaffold. Spin–phonon coupling introduces both energy relaxation and phase decoherence channels, limiting coherence times that are critical for quantum information processing and high-fidelity readout. This interaction also underpins phenomena in magnetic resonance imaging, optoelectronics, and quantum sensing, underscoring the need for precise theoretical frameworks to capture spin relaxation processes at the molecular scale.

\begin{figure}[tb]
 \includegraphics[height=6cm]{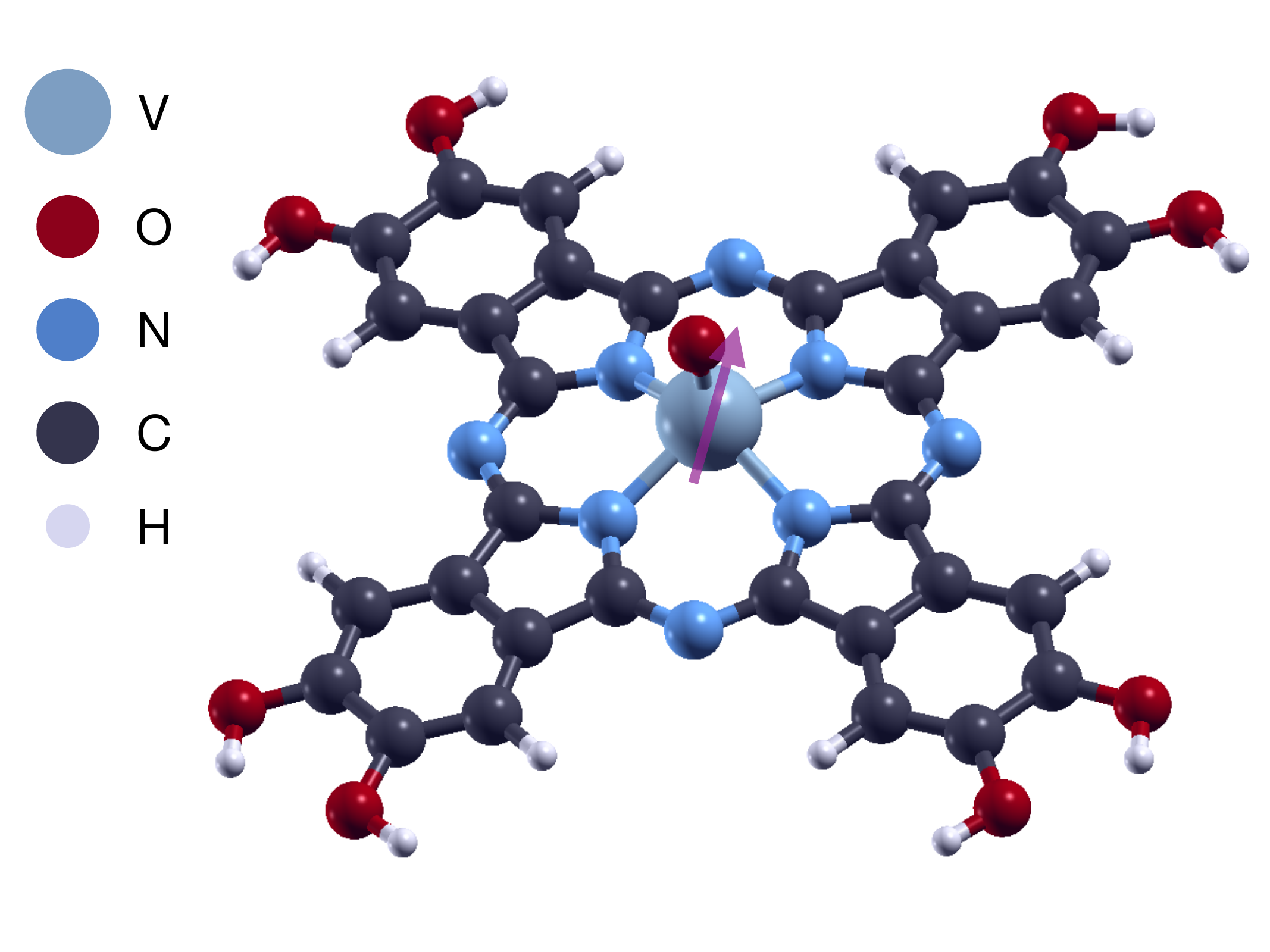}
    \caption{Molecular structure of \ce{VOPc(OH)8}, a VOPc derivative that is ivestigated in this work, with vanadium, oxygen, nitrogen, carbon, and hydrogen atoms shown in sky blue, red, blue, black, light gray, respectively. The spin-qubit electron is schematically shown with a purple arrow.}
    \label{fig:molecule}
\end{figure}

Vanadyl phthalocyanine (VOPc) has emerged as a prominent candidate for solid-state quantum technologies, including room-temperature quantum computing and molecular spin-based devices, owing to its long electron spin coherence times and structural robustness~\cite{Atzori:2016aa, Bonizzoni:2017aa, Malavolti:2018aa, Aziz:2011aa}. Fig.~\ref{fig:molecule} shows a single \ce{VOPc} molecule. The $d$-orbital of the \ce{V^{+4}} center hosts a single electron in a $d^1$ configuration responsible for the qubit character of this molecule. First-principles calculations, particularly those based on density functional theory (DFT), have become essential for investigating the electronic and magnetic properties of molecular magnets~\cite{ja061798aa, Timco:2009tc, Chilton:2013aa, Chilton2015, Reta:2021aa}. These methods enable the study of spin–phonon dynamics without relying on phenomenological parameters, thereby aiding the rational design of materials with tailored properties. However, challenges persist in accurately modeling these processes due to the high computational cost and the complexity of spin–phonon interactions, which involve a large number of vibrational degrees of freedom. For instance, a computational bottleneck appears due to the fact that the minimum number of separate DFT simulations required for the evaluation of second order spin-environment coupling is proportional to the square of total number of atoms in the system. Machine learning approaches~\cite{ML0} are capable of significant reduction in computational resources requirement~\cite{ Lunghi-single-ion-magnets, Lunghi2017}, but the predicted characteristic times of the relaxation processes were several orders of magnitude away from the experimental data. Therefore, to achieve accurate results, a full set of  calculations is necessary.~\cite{Lunghi_direct_raman_full}

Highlighting these shortcomings underscores the value of exploring alternative first-principles approaches in the theoretical and computational literature that focus on phonon-assisted spin relaxation modeling. Lunghi \textit{et al.} recently employed a first-principles method to model the spin dynamics of \ce{VO(acac)2}, focusing on first-order spin–phonon couplings through phonon-assisted modulation of the Zeeman and dipolar Hamiltonians~\cite{Lunghi2017}. This work supports the use of low-order Taylor expansions within Redfield theory for describing spin relaxation. However, even when modeling the full crystal environment and including phonons across the entire Brillouin zone using \(3 \times 3 \times 3\) supercells (containing 1620 atoms) at the DFT level, the resulting spin-relaxation times remained orders of magnitude away from experimental measurements~\cite{sciadv2019}. Similarly, other first-principles studies incorporating Raman and Orbach mechanisms for phonon-assisted relaxation in single-molecule magnets and qubits have captured experimental trends, but only within a few orders of magnitude~\cite{Lungi-direc_and_raman-VOacac, Lunghi-single-ion-magnets, Lunghi2017}. 
Despite these quantitative discrepancies, first-principles models remain invaluable for their ability to reproduce the qualitative features of phonon-assisted spin relaxation without resorting to empirical parameters. As such, they provide powerful tools for probing the microscopic origins of spin decoherence and for guiding the design of molecular architectures with enhanced quantum coherence.

Various phenomenological methods have been employed to deduce the functional form of the \(T_1\) variation as a function of magnetic field and temperature by fitting curves to experimental data ~\cite{atzori-t1, Santanni:2021ts}.
Additionally, researchers have used ligand field methods to systematically analyze the various phenomenological methods that have been employed to deduce the functional form of the \( T_1 \) dependence on magnetic field and temperature by fitting analytical expressions to experimental data~\cite{atzori-t1, Santanni:2021ts}. Additionally, ligand field models have been used to systematically analyse the contribution of individual phonon modes to spin relaxation via both direct and Raman processes. These analyses aim to correlate specific phonon symmetries with their influence on spin dynamics. However, such approaches typically require parameter adjustment to reproduce experimental \( T_1 \) values quantitatively~\cite{Ryan2021jacs}.

A notable semi-empirical effort has been presented by Aruachan \textit{et al.}, who constructed a parameterized Redfield quantum master equation to describe the interaction of molecular spin qubits with both lattice phonons and surrounding electron/nuclear spin baths~\cite{Aruachan}. Building on the Haken–Strobl theory~\cite{Capek1985}, this approach treats system–reservoir interactions as stochastic fluctuations in the system Hamiltonian. In this framework, spin–lattice and spin–spin couplings are modelled by a fluctuating gyromagnetic tensor and a fluctuating local magnetic field, respectively. The master equation is then derived to first order in these fluctuations. This semi-empirical approach introduces fitting parameters into the bath spectral densities of the Redfield tensor, which must be calibrated against experimental measurements.

While such models can yield excellent quantitative and qualitative agreement with experimental \( T_1 \) and \( T_2 \) data, they ultimately depend on prior experimental input for calibration and thus lack predictive generality.

In this work, we present a fully first-principles treatment of spin relaxation in a prototypical \ce{VOPc} derivative, \ce{VOPc(OH)8}, by directly computing the tensorial spin–phonon couplings and incorporating both linear and quadratic terms in the spin-vibrational Hamiltonian. Using a combination of DFT, numerical derivative techniques, and open quantum systems theory, we derive effective Lindblad and Redfield master equations that capture both direct one-phonon processes and higher-order Raman-type decoherence pathways. Importantly, our results reveal how a small number of low-frequency vibrational modes dominate spin relaxation, even in the absence of fitted parameters.

The structure of the paper is as follows: Section~\ref{sec:II} presents the spin–vibrational Hamiltonian and the derivation of the effective dissipators and concludes with an outline of our computational methodology.
Section~\ref{sec:III} presents our numerical results along with a direct 
comparison against recent experimental data. Section~\ref{sec:IV} concludes with a discussion of future directions for quantum coherence engineering in molecular systems.

\section{Spin–Phonon Relaxation Framework}\label{sec:II}

The objective of this section is to formulate a microscopic model of spin relaxation in \ce{VOPc(OH)8} based on mode-resolved spin--vibrational interactions. We begin by introducing the general form of the spin--vibrational Hamiltonian and derive an effective quantum master equation for the spin degrees of freedom by adiabatically eliminating the molecular vibrations. The resulting Lindblad and Redfield-type equations allow us to identify the dominant contributions to spin relaxation and dephasing, including both first-order and second-order (Raman-like) processes. Following this, we describe how the spin--vibrational coupling tensors appearing in the model are extracted from density functional theory (DFT) calculations, enabling a fully \textit{ab initio} prediction of spin coherence times without adjustable parameters.
\begin{widetext}

\subsection{Spin--Vibrational Hamiltonian}\label{sec:II.A}
Both single-molecule magnets (SMMs) and metal--organic frameworks (MOFs) typically feature one or a few central electronic spins embedded in environments of nuclear spins and molecular vibrations, as illustrated in Figure~\ref{fig:molecule}.
In transition-metal systems, the spin Hamiltonian is often dominated by the Zeeman interaction. Neglecting hyperfine and dipole--dipole contributions, the spin Hamiltonian is given by
\begin{equation}
\label{eq:zeeman}
    H_s = \beta \, \vec{S}   \mathbf{g}   \vec{B} = \sum_{\alpha=1}^{n_s} g_\alpha \hat{S}_\alpha,
\end{equation}
where \( \beta = \mu_B/\hbar \) is the ratio of the Bohr magneton to the reduced Planck constant, \( \vec{S} = (\hat{S}_x, \hat{S}_y, \hat{S}_z) \) is the spin operator vector, \( \vec{B} = (B_x, B_y, B_z)^T \) denotes the applied magnetic field, and \( \mathbf{g} \) is the $g$-tensor characterizing the coupling between the spin and the external field. Additionally, to keep the further derivations clean, we combine Bohr magneton $\beta$, the components of magnetic field $\vec B$ and g-tensor \( \mathbf{g} \) into single variable $g$.

We describe the vibrational degrees of freedom using a harmonic oscillator Hamiltonian:
\begin{equation}
    H_b = \sum_{k=1}^{n_q} \frac{1}{2} \left( \hat{p}_k^2 + \omega_k^2 \hat{x}_k^2 \right),
\end{equation}
where \( \hat{x}_k \) and \( \hat{p}_k \) denote the mass-weighted normal coordinates and their conjugate momenta, respectively, and the summation runs over the \( n_q \) vibrational modes of the molecule.

To capture how vibrational motion influences spin dynamics, we expand the spin Hamiltonian about the equilibrium geometry in terms of normal mode displacements and retain terms through second order:
\begin{equation}
    H = \sum_{\alpha=1}^{n_s} h_\alpha \hat{S}_\alpha + 
    \sum_{k=1}^{n_q} \frac{1}{2} \left( \hat{p}_k^2 + \omega_k^2 \hat{x}_k^2 \right) + \sum_{\alpha=1}^{n_s} \hat{S}_\alpha \left[ 
    \sum_{k=1}^{n_q} g_{\alpha k} \hat{x}_k + \sum_{k,k'=1}^{n_q} g^{(2)}_{\alpha k k'} \hat{x}_k \hat{x}_{k'} + \cdots
    \right]
\end{equation}
The first two terms describe the uncoupled spin and vibrational dynamics. The third and fourth terms account for spin--vibration couplings via mode-dependent modulations of the $g$-tensor. These couplings are obtained by expanding the $g$-tensor in a Taylor series around the equilibrium geometry and computing its derivatives along each normal mode direction. We evaluate these derivatives numerically using a finite-difference scheme, which requires computing the $g$-tensor for each distorted geometry. While conceptually straightforward, this procedure is computationally intensive and must be carried out separately for each system.  In terms of notation, we shall use lower-case Greek letters to denote the spin variable and $q$ or $k$ to denote normal coordinates. 
Unless specified otherwise, we take the laboratory $z$-axis defined by the external magnetic field \( \vec{B} = (0, 0, B_z)^T \) as the axis of quantization for the spin. 

We next derive expressions for spin relaxation and dephasing rates for a two level system $\left(\hat{S} = \frac{1}{2} \left\{\sigma_x, \sigma_y, \sigma_z\right\}\right)$ based on the general form of the spin--vibrational Hamiltonian. The specific coupling tensors that enter these expressions will be computed from first-principles electronic structure methods, as detailed in the following subsection. Under the assumptions that the spin--normal mode couplings \( g_{\alpha q} \) are weak, the mode frequencies \( \omega_q \) are off-resonant with the spin precession frequency \( \Omega \), and the normal modes relax rapidly due to environmental damping, we adiabatically eliminate the normal modes using the Born--Markov approximation. The resulting effective master equation for the reduced spin density matrix \( \rho_s(t) \) takes the Lindblad form:
\begin{equation}
    \frac{d\rho_s}{dt} = -\frac{i\Omega}{2} [\sigma_z, \rho_s] + \sum_{q=1}^N \Gamma_q \left( B_q \rho_s B_q^\dagger - \frac{1}{2} \{ B_q^\dagger B_q, \rho_s \} \right),
\end{equation}
where the effective jump operators \( B_q \) are given by
\begin{equation}
    B_q = \sum_{\alpha = x,y,z} g_{\alpha q} \sigma_\alpha,
\end{equation}
and the associated relaxation rates are
\begin{equation}
    \label{eq:first_broadening}
    \Gamma_q = \frac{4 \gamma_q}{\gamma_q^2 + 4 \omega_q^2} \left(n_q^{\mathrm{th}} + \frac{1}{2} \right),
\end{equation}
where $\gamma_q$, $\omega_q$, and $n_q^{th}$  are the relaxation rate, 
frequency, and thermal population of mode $q$, respectively.  Of these, the
$\gamma_q$'s are sole adjustable 
parameters of our model.

Expanding the dissipators in terms of spin components, we obtain
\begin{equation}
    \frac{d\rho_s}{dt} = -\frac{i\Omega}{2} [\sigma_z, \rho_s] + \sum_{\alpha,\alpha' = x,y,z} \Lambda_{\alpha\alpha'} \left( \sigma_\alpha \rho_s \sigma_{\alpha'}^\dagger - \frac{1}{2} \{ \sigma_{\alpha'}^\dagger \sigma_\alpha, \rho_s \} \right),
\end{equation}
with relaxation tensor
\begin{equation}
    \label{dtensor_first}
    \Lambda_{\alpha\alpha'} = \sum_{q=1}^{N} \Gamma_q g_{\alpha q} g_{\alpha'q}.
\end{equation}
\noindent
\( \Lambda \) is real, symmetric, and positive semi-definite. Its diagonalization yields the principal relaxation axes and rates. The relaxation times along and perpendicular to a quantization axis \( \hat{n} \) are given by
\begin{align}
\label{eq:tensor_projection_1}
    \frac{1}{T_1} &= 2 \hat{n}^T \Lambda \hat{n}, \\
    \frac{1}{T_2} &= \mathrm{Tr}[\Lambda] - \hat{n}^T \Lambda \hat{n} = \mathrm{Tr}[\Lambda] - \frac{1}{2T_1}.
\end{align}
For \( \hat{n} = \hat{z} \), these reduce to the familiar expressions:
\begin{align}
\label{eq:tensor_projection_2}
    \frac{1}{T_1} &= 2\Lambda_{zz}, \\ 
    \frac{1}{T_2} &= \Lambda_{xx} + \Lambda_{yy}.
\end{align}
relating the population relaxation rate $1/T_1$ to the phase relaxation 
$1/T_2$ rate.  (Note that this is only the dephasing due to relaxation, the total dephasing rate also includes a pure dephasing 
component, which generally dominates the dephasing in a molecular system.)
Diagonalizing \( \Lambda \) as \( \Lambda = \sum_{k=1}^3 \lambda_k \vec{v}_k \vec{v}_k^T \) with eigenvalues \( \lambda_k \geq 0 \) and orthonormal eigenvectors \( \vec{v}_k \), we identify the natural axes and rates of dissipation. This provides a transparent characterization of spin decoherence due to anisotropic, mode-structured bosonic baths.

\subsection{Higher-Order Contributions and Raman Relaxation Pathways}
\label{sec:II.B}
Spin relaxation in molecular systems is not limited to direct one-phonon processes. Second-order relaxation pathways, commonly referred to as Raman processes, arise from virtual transitions mediated by either linear or quadratic spin--vibrational couplings. These higher-order effects can dominate spin relaxation, particularly when quadratic coupling terms are significant, as is the case for \ce{VOPc(OH)8}.

The second-order transition rate arises from fourth-order perturbation theory and involves the four-point correlation function of vibrational coordinates. Assuming the vibrational bath is Gaussian and stationary, and applying Wick's theorem, we obtain
\begin{equation}
    \Gamma_{\alpha\alpha'}^{(2)} = \frac{2}{\hbar^4} \sum_q g_{\alpha q}^2 g_{\alpha'q}^2 \int_0^\infty dt \, e^{i\Omega t} \langle  x_q(t)  x_q(0) \rangle^2,
\end{equation}
where the two-point correlation function of the harmonic bath is 
\begin{equation}
    \langle  x_q(t)  x_q(0) \rangle = (2n_q + 1) \cos(\omega_q t).
\end{equation}

Using the identity \( \cos^2(\omega_q t) = \tfrac{1}{2} + \tfrac{1}{2} \cos(2\omega_q t) \), we obtain a regularized expression for the Raman contribution to the relaxation tensor:
\begin{equation}
    \Lambda_{\alpha\alpha'}^{(2)} = \frac{1}{\pi \hbar^4} \sum_q g_{\alpha q}^2 g_{\alpha'q}^2 (2n_q + 1)^2    \frac{\lambda_q}{(\Omega - 2\omega_q)^2 + \lambda_q^2},
\end{equation}
where \( \lambda_q \) is a phenomenological linewidth parameter that accounts for vibrational dephasing. In the high-temperature limit \( k_B T \gg \hbar \omega_q \), the expression simplifies to
\begin{equation}
    \Lambda_{\alpha\alpha'}^{(2)}(T \gg \hbar\omega_q / k_B) =  (k_B T)^2\frac{8}{\pi \hbar^6}   \sum_q \frac{g_{\alpha q}^2 g_{\alpha' q}^2}{\omega_q^4}   \frac{\lambda_q}{(\Omega - 2\omega_q)^2 + \lambda_q^2},
\end{equation}
and we recover the characteristic 
$1/T_1 \propto T^2$ scaling of the relaxation 
rate due to second-order processes. 

An analogous second-order contribution arises from quadratic spin--vibrational couplings of the form
\begin{equation}
    H^{(2)}_{\text{int}} = \sum_{\alpha} \sigma_\alpha \sum_q g^{(2)}_{\alpha qq}  x_q^2,
\end{equation}
which leads to additional Raman-type terms in the master equation. Assuming Gaussian statistics for \(  x_q \), Wick's theorem gives
\begin{equation}
    \langle  x_q^2(t)  x_q^2(0) \rangle = (2n_q + 1)^2 \left[ 2\cos^2(\omega_q t) + 1 \right].
\end{equation}
Evaluating the Fourier transform as before, we obtain the total second-order contribution to the relaxation tensor:
\begin{equation}
    \label{dtensor_second}
    \Lambda_{\alpha\alpha'}^{(2,\text{total})} = \sum_q \frac{(2n_q + 1)^2}{\pi}   \frac{\lambda_q}{(\Omega - 2\omega_q)^2 + \lambda_q^2} \left[ \left( \frac{g_{\alpha q}}{\omega_q} \right)^2 \left( \frac{g_{\alpha'q}}{\omega_q} \right)^2 + g^{(2)}_{\alpha qq} g^{(2)}_{\alpha' qq} \right].
\end{equation}
These processes provide substantial contributions to spin relaxation and can even dominate over first-order pathways, depending on the mode structure and coupling strength. The relaxation tensor is given by
\begin{equation}
    \Lambda_{\alpha\alpha'}^{\text{total}} = \Lambda_{\alpha \alpha'}^{(1)} + \Lambda_{\alpha\alpha'}^{(2)},
\end{equation}
where \( \Lambda_{\alpha}^{(1)} \) includes contributions from direct one-phonon processes and \( \Lambda_{\alpha\alpha'}^{(2)} \) includes both virtual transitions and quadratic couplings.

Spin relaxation may also proceed via \emph{virtual two-phonon transitions}, where the spin-flip frequency \( \Omega \) is resonant with twice the vibrational frequency of a mode. These processes correspond to resonant two-phonon scattering and are second order in the spin--vibrational coupling. We refer to these as \emph{resonant-Raman} processes to distinguish them from non-resonant Raman relaxation or Orbach-type spin flips involving real excited states.

These processes can be modeled within the Lindblad formalism by including spin operators coupled to quadratic combinations of vibrational coordinates. We consider a second-order interaction Hamiltonian of the form:
\begin{equation}
H_{\mathrm{int}}^{(2)} = \sum_{\alpha} \sigma_\alpha \sum_{q, q'} g^{(2)}_{\alpha qq'}  x_q  x_{q'},
\end{equation}
where \(  x_q \) is the displacement operator for mode \( q \), and \( g^{(2)}_{\alpha qq'} = \partial^2 H_{\text{spin}} / \partial  x_q \partial  x_{q'} \) are the second-order spin--vibrational coupling coefficients for spin component \( \sigma_\alpha \). 
Retaining only the diagonal term $q=q^{'}$ gives the Lindblad-type master equation for the spin:
\begin{equation}
\left.\frac{d\rho_s}{dt}\right|_{\mathrm{Raman}} =  \sum_{\alpha} \Gamma_\alpha^{(2)} \left( \sigma_\alpha \rho_s \sigma_\alpha - \frac{1}{2} \{ \sigma_\alpha^2, \rho_s \} \right),
\end{equation}
with relaxation rates given by
\begin{equation}
\Gamma_\alpha^{(2)} = \int_0^\infty dt\, e^{i\Omega t} \left\langle B_\alpha(t) B_\alpha (0) \right\rangle, \qquad B_\alpha = \sum_q g^{(2)}_{\alpha qq}  x_q^2.
\end{equation}

Using Wick’s theorem, the four-point correlation function simplifies to
\begin{equation}
\left\langle  x_q^2(t)  x_q^2(0) \right\rangle = (2n_q + 1)^2 \left[ 2 \cos^2(\omega_q t) + 1 \right],
\end{equation}
where \( n_q = (e^{\hbar \omega_q / k_B T} - 1)^{-1} \) is the thermal occupation number. Evaluating the time integral yields
\begin{equation}
\Gamma_\alpha^{(2)} = \sum_q \left( g^{(2)}_{\alpha qq} \right)^2 (2n_q + 1)^2 \left[ \pi \delta(\Omega) + \frac{\pi}{2} \delta(\Omega - 2\omega_q) \right].
\label{eq:25}
\end{equation}
The first term corresponds to elastic dephasing; the second describes inelastic resonant-Raman relaxation.
In the high-temperature limit, \( (2n_q + 1)^2 \approx \left( \frac{2k_B T}{\hbar \omega_q} \right)^2 \), leading to:
\begin{equation}
\Gamma_\alpha^{(2)} \sim T^2 \sum_q \left(\frac{ g^{(2)}_{\alpha qq} }{\omega_q}\right)^2 \delta(\Omega - 2\omega_q).
\end{equation}

Including both linear two-phonon contributions and second-derivative couplings, and regularizing the \( \delta \)-function as a Lorentzian of width \( \lambda_q \), we obtain:
\begin{equation}
%\boxed{
\Gamma_\alpha^{(2,\mathrm{total})} =  \frac{1}{\pi}\sum_q (2n_q + 1)^2     \frac{\lambda_q}{(\Omega - 2\omega_q)^2 + \lambda_q^2}   \left[ \left(\frac{g_{\alpha q}}{\omega_q}\right)^4 + \left( g^{(2)}_{\alpha qq} \right)^2 \right]
\label{eq:lindblad_final}
%}
\end{equation}
This expression provides a unified treatment of resonant two-phonon spin relaxation via both first- and second-order spin--vibrational couplings.
In practice, we extend this 
to include both on- and off-diagonal 
contributions to the $g$-tensor Hessian, $g^{(2)}_{\alpha q q'}$. We also allow for simultaneous excitation and di-excitation of two separate normal modes, which transforms Eq.~\ref{eq:25} to:
\begin{align}
\label{eq:second_broadening}
\Gamma_\alpha^{(2)} =\frac{1}{4} \sum_q \left( g^{(2)}_{\alpha qq} \right)^2  
[
&\delta(\Omega - \omega_{q} - \omega_{q'})n_{q}n_{q'} +
\delta(\Omega + \omega_{q} + \omega_{q'})(n_{q} + 1)(n_{q'} + 1) \nonumber \\
+ &\delta(\Omega + \omega_{q} - \omega_{q'}) (n_{q} + 1) n_{q'} +
\delta(\Omega - \omega_{q} + \omega_{q'}) n_{q} (n_{q'} + 1)
]
\end{align}
As we shall discuss below, this gives important insight into 
which modes or combinations of normal modes of the molecular system
contribute dominantly to the relaxation and decoherence rates.

\end{widetext}

\subsection{First-Principles Evaluation of the Relaxation Tensor
}\label{sec:II.C}

The relaxation tensor \( \Lambda_{\alpha\alpha'} \) introduced above encodes the anisotropic coupling between the spin and the thermally populated vibrational modes of the molecule. To evaluate this tensor from first principles, we compute the mode-resolved spin--vibrational coupling coefficients that enter both the first- and second-order relaxation pathways. Specifically, these couplings arise from the sensitivity of the $g$-tensor to normal mode displacements.

We define the first-order coupling coefficients as
\begin{equation}
    \label{eq:g_finite_diff_1}
    g_{\alpha k} = \left. \frac{\partial g_{\alpha}}{\partial x_k} \right|_{x = 0},
\end{equation}
where \( x_k \) denotes the mass-weighted normal coordinate of mode \( k \), and \( g_\alpha \) refers to the Cartesian components of the $g$-tensor (\( \alpha = x, y, z \)). These derivatives are evaluated numerically using a central finite-difference scheme:
\begin{equation}
\label{eq:g_finite_diff_2}
    g_{\alpha k} \approx \frac{g_\alpha(x_k + \delta) - g_\alpha(x_k - \delta)}{2\delta},
\end{equation}
where \( \delta \) is a small displacement amplitude along mode \( k \). Analogous expressions are used to compute second-order derivatives:
\begin{equation}
\label{eq:g_finite_diff_3}
    g^{(2)}_{\alpha kk'} = \left. \frac{\partial^2 g_\alpha}{\partial x_k \partial x_{k'}} \right|_{x = 0},
\end{equation}
with the diagonal terms \( k = k' \) contributing directly to Raman-type two-phonon relaxation.

All derivatives are taken with respect to the mass-weighted normal coordinates, which ensures consistent units and scaling in the vibrational Hamiltonian. In practice, these derivatives are obtained by distorting the molecular geometry along each normal mode direction, recalculating the $g$-tensor at each displaced geometry, and performing numerical differentiation.
Equivalently, displacements along the normal mode coordinates can be expressed in terms of the Cartesian displacements via chain rule: 
\begin{equation}
\label{eq:g_finite_diff_4}
    \frac{\partial}{\partial x_k} = 
    \sum_{i}
    \frac{\partial}{\partial R_i}
    \frac{\partial R_i}{\partial x_k} = 
    \sum_{i}
    \sqrt{\frac{\hbar}{\omega_k}}L_{ik}
    \frac{\partial}{\partial R_i} ,
\end{equation}
\begin{equation}
\label{eq:g_finite_diff_5}
    \frac{\partial^2}{\partial x_k \partial x_{k^{'}}} = 
    \sum_{i}\sum_{j}
    \sqrt{\frac{\hbar}{\omega_k^{\phantom{'}}}}
    \sqrt{\frac{\hbar}{\omega_{k^{'}}}}
    L_{ik^{\phantom{'}}}L_{jk^{'}}
    \frac{\partial^2}{\partial R_i \partial R_j} ,
\end{equation}
where $R_i=\{x_i, y_i, z_i\}$ represents Cartesian coordinates of $i$-th atom, $\omega_k$ is the $k$-th normal mode frequency and $L$ is a mass-weighted matrix of normal modes eigenvectors. These quantities are the direct result of the diagonalization of mass-weighed Hessian and are readably accessible from the vibrational properties calculation performed in a first principles software package. 

In this work all electronic structure calculations were carried out using the \textsc{ORCA} quantum chemistry package (version 6.0)~\cite{orca}. The molecular geometry of \ce{VOPc(OH)8} was optimized using unrestricted open-shell density functional theory (DFT) with the PBE exchange--correlation functional~\cite{PBE}. Tight convergence thresholds were applied via the \texttt{TIGHTOPT} keyword for geometry optimization (energy convergence threshold of $10^{-6}$ Hartree) and \texttt{TIGHTSCF} for electronic self-consistency (threshold of $10^{-8}$ Hartree).
A segmented basis set scheme was employed to balance accuracy and computational efficiency: the def2-TZVP basis set was used for vanadium and oxygen atoms, def2-SVP for carbon, nitrogen, and hydrogen, and the def2-TZVP/C auxiliary basis set for Coulomb fitting~\cite{weigned2005basis, weigned2006coulomb}. No solvent models was used in this work; calculations were performed on the system in vacuum. Normal mode frequencies and displacements were obtained via harmonic vibrational analysis at the optimized geometry. The resulting modes were mass-weighted and used to define distortion directions for finite-difference evaluation of $g$-tensor derivatives.

First- and second-order derivatives of the $g$-tensor with respect to each normal mode coordinate were computed numerically using central finite differences. Each mode was distorted by \( \pm \delta \) along the mass-weighted normal coordinate, and the $g$-tensor was recalculated at each displaced geometry. The derivatives were then assembled into the spin--vibrational coupling tensors \( g_{\alpha k} \) and \( g^{(2)}_{\alpha kk'} \), which enter the relaxation tensor expressions described above.

\begin{figure*}[htb]
    \centering
    \includegraphics[width=\linewidth]{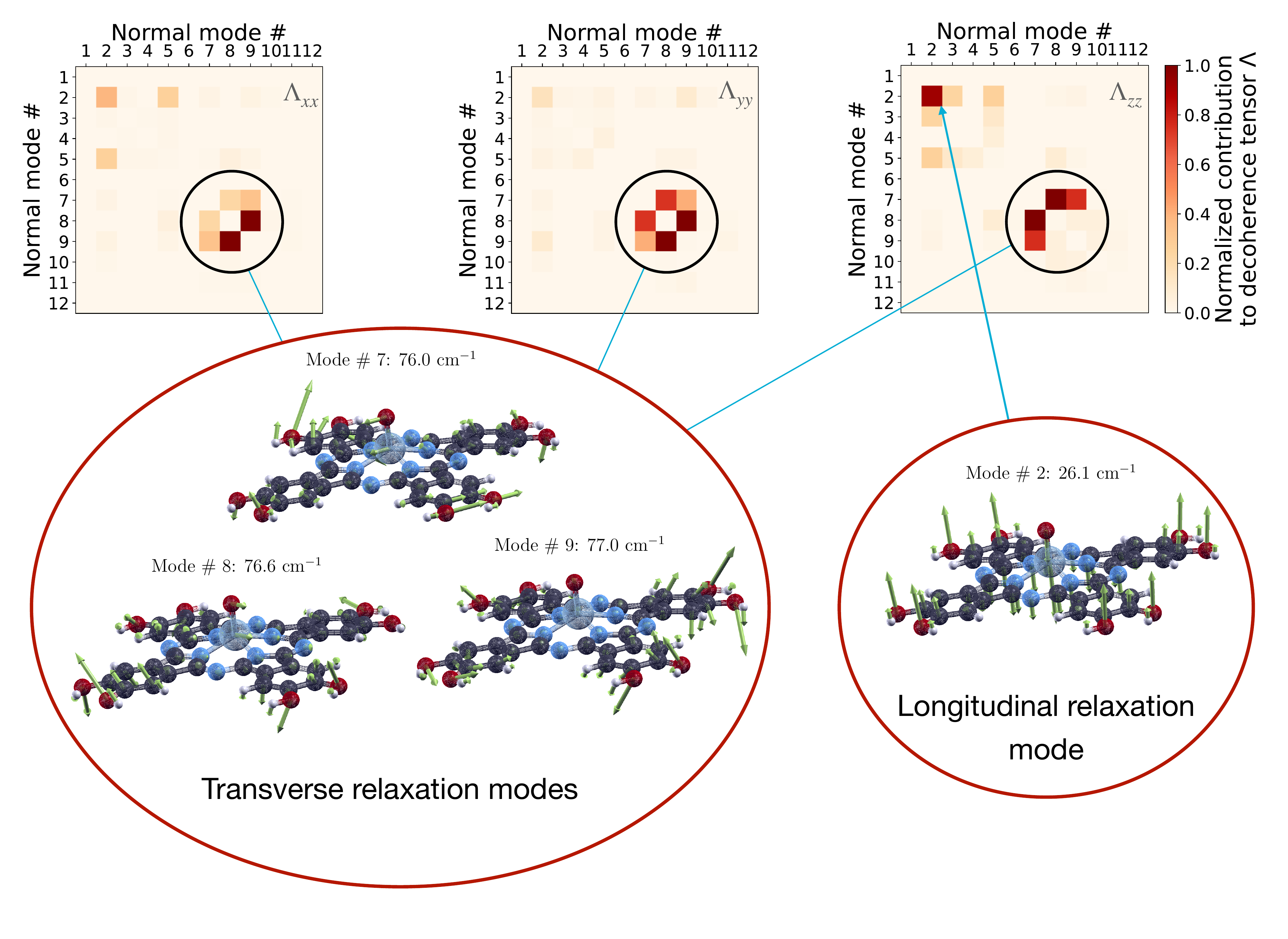}
    \caption{Normal modes that define the relaxation tensor. (Top) Heat maps that show the normalized contribution from the first 12 normal modes to the second-order relaxation tensor $\Lambda^{(2)}$, which includes the Lorentzian line shape broadening centered at the spin transition frequency $\Omega=1.2$ cm$^{-1}$ with linewidth of $2$ cm$^{-1}$. Vibrational modes with numbers 7, 8, and 9 contribute to the components of $\Lambda^{(2)}$ which define both transversal and longitudinal relaxation, while mode number 2 uniquely influences the the components of $\Lambda^{(2)}$ responsible for the transverse relaxation. (Bottom) Directional force plots of the highlighted normal modes. For this figure only twelve lowest lying modes of the vibrational spectrum were considered, as the remaining higher frequency modes have a quantitatively insignificant contribution to the decoherence tensor and hence were omitted.}
    \label{fig:heat_modes}
\end{figure*}

This \textit{ab initio} protocol yields a complete parameterization of the spin--vibrational Hamiltonian and associated relaxation tensor without invoking any empirical fitting. It enables mode-resolved analysis of spin relaxation pathways rooted in the electronic and vibrational structure of the molecule.
This procedure is computationally demanding, as it requires a full electronic structure calculation for each distorted configuration. However, it yields direct access to the spin--vibrational coupling landscape without introducing empirical parameters or model assumptions. The resulting tensors \( g_{\alpha k} \) and \( g^{(2)}_{\alpha kk'} \) are then used to evaluate the relaxation tensor \( \Lambda_{ij} \) and the associated relaxation times \( T_1 \) and \( T_2 \).

\subsection{Comparison to Redfield theory
}\label{sec:II.D}

We compare the numerical results of our theory with a different Markovian master equation approach, as derived within the framework of Redfield ~\cite{REDFIELD_main} theory. An excellent derivation of an extension of Redfield theory to higher orders of spin-phonon coupling can be found in the work by Lunghi and Sanvito~\cite{Lungi-direc_and_raman-VOacac}, which we adopt here. The characteristic relaxation times were obtained by numerically solving Bloch-Redfield master equation using Qutip: Quantum Toolbox in Python~\cite{QUTIP} version 4.7.5 compatible with SciPy~\cite{SCIPY} python package version 1.12.0. Numerical integration was performed by \textit{brmesolve()} routine with 10000 time steps for the total of $10^4$ microsecond for $T_1$ and 20000 time steps for the total of 10 microseconds for $T_2$ to insure that the spin populations are relaxed back to the ground state while maintaining numerical stability.

\section{Results and Discussion}\label{sec:III}

For \ce{VOPc(OH)8}, we computed the $g$-tensor in the molecular frame:
\begin{equation}
\bm{\mathsf{g}} =
\begin{pmatrix}
1.981 & \phantom{-}3.923 \times 10^{-3} & \phantom{-}2.134 \times 10^{-3} \\
\phantom{-}3.897 \times 10^{-3} & 1.989 & -8.711 \times 10^{-4} \\
\phantom{-}2.119 \times 10^{-3} & -8.722 \times 10^{-4} & 1.990
\label{eq:g_baseline}
\end{pmatrix}.
\end{equation}
The tensor is not diagonal because it is defined in the coordinate frame specified by the molecular principal rotational axes. This poses no fundamental difficulty, as it can be diagonalized by rotating to the laboratory frame aligned with the applied magnetic field. In this rotated frame, the principal values are found to be \((g_x, g_y, g_z) = (1.979, 1.991, 1.991)\).
These computed values differ only slightly from the experimental values \((g_x, g_y, g_z) = (1.966, 1.989, 1.989)\) reported in Ref.~\citenum{Ryan2021jacs} for \ce{VOPc}. No empirical adjustments were made to the PBE exchange–correlation functional to enforce agreement with experiment; the results are entirely first-principles.

The corresponding optimized molecular geometry was subsequently perturbed to obtain the first and second order derivatives of g-tensor with respect to normal mode dimensionless coordinates $x_k$ according to Eq. \ref{eq:g_finite_diff_1} -- \ref{eq:g_finite_diff_5}. The magnitude of the perturbation of $0.01$ Angstrom ensured that the change in g-tensor components due to the geometrical variation were at least two orders of magnitude larger than the numerical noise.  We include our calculated $g$ and $g^{(2)}$ in the Supplementary Information. 

We proceed with the evaluation of the decoherence tensor $\Lambda$, which bridges the DFT approach with Lindblad theory and plays the defining role in the relaxation dynamics. Following Eq.~\ref{dtensor_first} and~\ref{dtensor_second}, for a given temperature the derivatives of g-tensor enter the expression for the decoherence tensor with relaxation rates $\Gamma$ acting as a filter for vibrational spectrum. Broadened by the Lorentzian lineshapes with linewidth $\lambda_q$ centered at the spin transition frequency $\Omega$, the rates for direct (Eq.~\ref{eq:first_broadening}) and Raman (Eq.~\ref{eq:second_broadening}) excitation channels favor low lying normal modes with frequencies below $400$ cm$^{-1}$ (Fig. S1). Such selective behavior is explained by two key factors. First, for common experimental magnetic fields ($\sim 1T$) the \ce{VOPc(OH)8} spin transition frequency is of the order of a few cm$^{-1}$ and is significantly smaller than the first normal mode frequency calculated to be $12.6$ cm$^{-1}$. Second, for a broad range of linewidths $\lambda_q\in[1, 100]$ cm$^{-1}$ the vibrational spectral range that contains majority of the contributions to both $\Lambda^{(1)}$ and $\Lambda^{(2)}$ remains unchanged (Fig. S2, S3). For molecular qubits of comparable configurations these modes have linewidths of $1-10$ cm $^{-1}$.~\cite{chilton_phonon_lifetimes} For the sake of clear presentation of the data, in this work we chose the width $\lambda_q$ of the Lorentzians to be constant among all the phonons and equal to $2$ cm$^{-1}$. 

Having established the precise numerical framework for the evaluation of decoherence tensor, we compare the absolute values of the elements of $\Lambda^{(1)}$ and $\Lambda^{(2)}$. We discover that the contributions originating from the second derivatives of g-tensor is on average two orders of magnitude higher than those from the first order g-tensor derivatives (Fig. S2, S3). This observation automatically renders the 4-th order term $\sim (g_{\alpha q})^4$ that apears in Eq.~\ref{eq:lindblad_final} negligible. Accordingly, we focus on second order derivatives of g-tensor and present a more detailed analysis of $\Lambda^{(2)}$ in Fig.~\ref{fig:heat_modes}, where we demonstrate that there is, in fact, four vibrational normal modes out of total 192 that majorly define the second order decoherence tensor. Modes 7, 8 and 9 have nearly identical frequencies and are dominated by the motions of oxygen atoms on the perimeter of the molecule, while mode 2 corresponds to symmetric out-of-plane motions. We note that while the temperature explicitly enters the expressions for the relaxation rates and hence the decoherence tensor, the main underlying quantity for this analysis is the second-order derivatives of g-tensor with respect to dimensionless normal mode coordinates, which do not depend on temperature.

\begin{figure*}[]
    %\centering
    \subfigure{\includegraphics[width=0.47\textwidth]{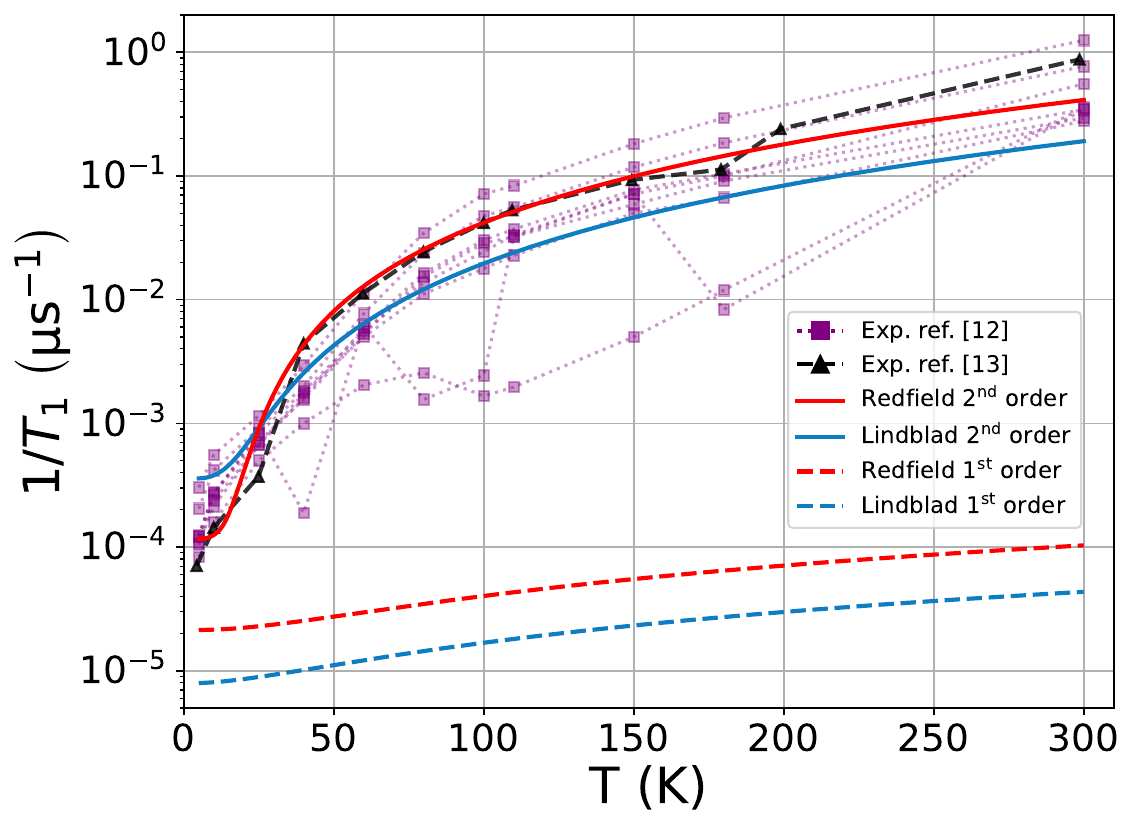}}
    \subfigure{\includegraphics[width=0.47\textwidth]{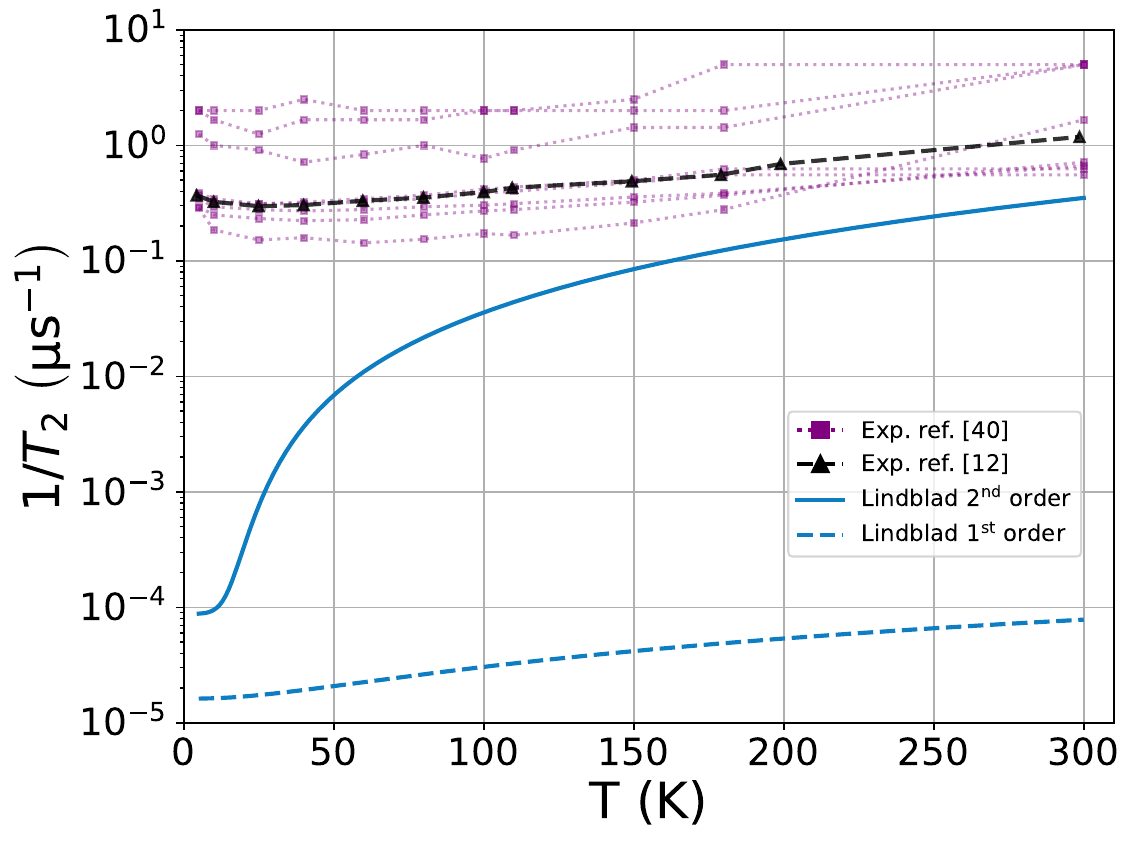}}
    \caption{The inverse of spin population relaxation time $1/T_1$ (left) and spin dephasing time $1/T_2$ (right) vs. temperature. Computational results, evaluated within the frameworks of first and second order Lindblad and Redfield theories, with external magnetic field of 1266 mT and lineshape broadening of 2 cm$^{-1}$, are shown in blue and red, correspondingly. Experimental data by Follmer \textit{et al.}~\cite{exp_main_hadt}, gathered within multiple measurements at magnetic fields between 300 and 400 mT (specifically 303 mT, 329 mT, 336 mT and 386 mT) and 1200 and 1300 mT (specifically 1198 mT, 1214 mT, 1218 mT and 1266 mT), shown in purple. Experimental data by Atzori \textit{et al.}~\cite{Atzori:2016aa}, measured at 345 mT, shown in black. }
    \label{fig:t1t2}
\end{figure*}

To thoroughly test our theory, we use Eq.~\ref{eq:lindblad_final} to calculate the relaxation time $T_1$ and dephasing time $T_2$ for \ce{VOPc(OH)8} molecule as a function of temperature. In Fig.~\ref{fig:t1t2} we compare our results to experimental measurements of Atzori \textit{et al.}~\cite{Atzori:2016aa} and Follmer \textit{et al.}~\cite{ exp_main_hadt}. We also include the calculation based on Redfield (see section~\ref{sec:II.D}) theory, where we used our calculated data for spin-phonon coupling terms to construct the Redfield tensor and numerically solve the master equation to obtain $T_1$. Our results based on Lindbald-like approach show an excellent agreement with experimental data and Redfield theory alike for the whole temperature range for $T_1$ and approach the experimental results at hight temperatures for $T_2$, while being purely analytical in nature. In general, the discrepancies can be attributed to the fact that here we only use $\Lambda_{zz}$ component of the decoherence tensor in our calculation, which we single out by performing a projection onto $z$-axis in going from Eq.~\ref{eq:tensor_projection_1} to Eq.~\ref{eq:tensor_projection_2}. However, the proximity of our $T_1$ estimation to Redfield theory result is not surprising since the $\Lambda_{zz}$, $\Lambda_{yy}$ and $\Lambda_{xx}$ are of the same order of magnitude (Fig. S3), and are expected to dominate the population relaxation dynamics for the two-level systems modeled by Zeeman Hamiltonian.~\cite{slichter2013principles} Expression for $T_2$, on the other hand, as formulated in Eq.~\ref{eq:tensor_projection_1}, does not take into account the pure dephasing effects, which especially affects the total dephasing time at lower temperatures.
 
 Reinforcing the conclusion of the analysis of decoherence tensor, the absolute values of $T_1$ and $T_2$ are completely defined by the contributions from the second derivatives of the g-tensor to $\Lambda^{(2)}$. The evolution of characteristic times with temperature, governed solely by the phonon population numbers combined with Lorentzian lineshapes (Eq.~\ref{eq:25}), is almost fully captured by $\Lambda^{(2)}$ as well, while first-order contributions play only a marginal role. At lower temperatures, the functional form  of temperature dependence is dictated by Bose-Einstein statistics for phonon pairs, while at higher temperatures ($T > 70$ K), the temperature scaling becomes classical and proportional to $T^2$. The impact of linewidth $\lambda_q$ is comparably small, mostly affecting the steepness of temperature curve at $T<50$ K. We note that at room temperatures and beyond, the higher order spin-phonon coupling terms may need to be included in order to fully capture how $T_1$ and $T_2$ evolve with temperature.

In the last step of our analysis, we discuss the influence of the external magnetic field $B$ on characteristic times in both the experimental measurements and our model. Follmer \textit{et al.} (purple color in Fig.\ref{fig:t1t2}) performed two distinct sets of measurements, taken at different external magnetic field ranges: between $300$ mT and $400$ mT, and between $1200$ mT and $1300$ mT. Despite this significant difference in the field, both the temperature dependence and the absolute values of experimental $T_1$ and $T_2$ are surprisingly close for both ranges, with data points often intercepting each other. In contrast, in our model the magnetic field enters as $B_z^2$ (Eq.~\ref{eq:zeeman} and ~\ref{eq:25} ) for $T_1$ and $T_2$ across the full temperature range, making the model highly sensitive to the variations in its magnitude. As a result, our theoretical model is capable to correctly predict the experimental data in only in one of the aforementioned ranges of $B$. Specifically, the calculations in this work were performed with magnetic field set to $1266$ mT.

\section{Conclusion}\label{sec:IV}
\label{sec:summary}
Our method efficiently explores spin--lattice relaxation processes with high accuracy. We applied this approach to investigate the spin dynamics in \ce{VOPc(OH)_8}, a promising building block for molecular magnets and filters. Leveraging this method, we have deepened our understanding of the physics underlying spin relaxation in molecular magnets by discovering that only a small, localized subset of vibrational normal modes plays an active role in the relaxation of the spin system. Second-order perturbation theory with respect to collective vibrational motions in the molecular bath is essential for correctly evaluating both the absolute values of the relaxation times $T_1$ and $T_2$ and their temperature dependence. While computationally demanding, calculating the second derivative of the $g$-tensor enabled accurate reproduction of the experimental temperature trend \( \sim T^2 \).

Our study demonstrates that spin--phonon relaxation in molecular qubits is governed by a highly mode-selective and symmetry-resolved structure, which can be captured quantitatively using a tensorial framework derived entirely from first principles. By combining vibrational analysis with $g$-tensor derivatives, we identify specific vibrational modes that dominate either longitudinal or transverse relaxation, with excellent agreement against experimental $T_1$ and $T_2$ measurements. The approach developed here provides a general and transferable strategy for computing relaxation tensors in molecular spin systems, enabling the predictive identification of decoherence pathways without empirical fitting. These results establish a foundation for engineering molecular environments that suppress spin relaxation through targeted vibrational control, and offer a new route to optimizing quantum coherence in chemically tunable systems.

\begin{acknowledgments}
    The research presented in this article was supported by the LANL LDRD program (number 20220047DR). 
LANL is operated by Triad National Security, LLC, for the National Nuclear Security Administration of the U.S. Department of Energy (contract no. 89233218CNA000001). The authors thank the LANL Institutional Computing (IC) program for access to HPC resources. 
The work at the University of Houston was funded in part by the National Science Foundation (CHE-2404788)  
and the Robert A. Welch Foundation (E-1337).
This work was performed, in part, at the Center for Integrated Nanotechnologies, an Office of Science User Facility operated for the U.S. Department of Energy (DOE) Office of Science by Los Alamos National Laboratory (Contract 89233218CNA000001) and Sandia National Laboratories (Contract DE-NA-0003525).
ERB also acknowledges support from the LANL Center for Nonlinear Studies (CNLS) through an Ulam Faculty Fellowship for 2022–2023.
\end{acknowledgments}

\section*{Data Availability Statement}
The data supporting this study's findings are available via the following GitHub repository \url{https://github.com/NosheenYounas/Spin-Phonon-Dynamics-1-2}

\section*{Author Contribution Statement}

    The authors confirm their contribution to the paper: study conception and design: ERB, YZ, and AP; data collection and simulations: NY and RD; analysis and interpretation of results: RD, NY, YZ, AP, and ERB; draft manuscript preparation: RD, ERB;
    development of adiabatic elimination approach: ERB.
    All authors reviewed the results and approved the final version of the manuscript.

\vskip2pc

\bibliography{}

\end{document}